\documentclass[12pt]{article}
\usepackage{epsfig}
\begin{document}                   
\title{Near-threshold $\phi$ meson production in proton--nucleus reactions and $\phi$ width in finite nuclei}
\author{E.Ya. Paryev\\
{\it Institute for Nuclear Research, Russian Academy of Sciences,}\\
{\it Moscow 117312, Russia}}

\renewcommand{\today}{}
\maketitle

\begin{abstract}
       In the framework of the nuclear spectral function approach for incoherent primary proton--nucleon and secondary
pion--nucleon production processes we study the inclusive $\phi$ meson production in the interaction of 2.83 GeV protons
with nuclei. In particular, the A- and momentum-dependences of the absolute and relative $\phi$ meson yields are investigated within the
different scenarios for its in-medium width. 
 Our model calculations take into account the acceptance
window of the ANKE facility used in a recent experiment performed at COSY.  They show that the 
pion--nucleon production channel contributes distinctly to the "low-momentum" 
$\phi$ creation in heavy nuclei in the chosen kinematics and, hence,
has to be taken into consideration on close examination of the dependences of the phi meson yields on the target mass number with
the aim to get information on its width in the medium. They also demonstrate that both the A-dependence of the relative $\phi$ meson
production cross section and momentum-dependence of the absolute $\phi$ meson yield 
at incident energy of interest are appreciably sensitive to the phi in-medium
width, which means that these observables can be useful to help determine the above width from the direct comparison
the results of our calculations with the future data from the respective ANKE-at-COSY experiment.
\end{abstract}
                                                  
\newpage

\section*{1 Introduction}

\hspace{1.5cm} Studies of modification of the $\phi$ meson spectral function at finite baryon density through its production and
decay in pion--nucleus [1], proton--nucleus [2--6] and photon--nucleus [7--10] reactions have received considerable interest
in recent years. Because, on the one hand, the $\phi$ meson (in contrast to the $\rho$ and $\omega$ mesons) does not overlap
with other light resonances in the mass spectrum and, on the other hand, the reactions on ordinary nuclei with elementary probes
are less complicated compared to heavy--ion collisions especially in the near-threshold energy regime where a small number of
possible channels for meson production contributes, one may hope to get from these studies a more clear precursor signals for
partial restoration of chiral symmetry--the fundamental symmetry of QCD in the limit of massless quarks--and thus to test this
prominent feature of QCD as well as to extract valuable information on the nucleon strangeness content [4, 7] and on the kaon
in-medium properties [1--4]. 
As is expected [11--14], the mass shift of the $\phi$ meson in nuclear matter is small, whereas its 
total in-medium width is appreciably increased compared to the vacuum value ($\Gamma_{\phi}=4.45$ MeV). 
The nontraditional possibility to learn about this width
has been considered in [2, 3, 6, 8--10]. As a measure for the in-medium broadening of the $\phi$ meson 
the A-dependence of its production cross section in nuclei in proton- and photon-induced reactions has been employed
in these works. The A-dependence is governed by the absorption of the $\phi$ meson flux in nuclear matter, which in turn is 
determined, in particular, by the phi in-medium width. The advantage of this method is that one can exploit the main decay channel
$\phi \to K^+K^-$, having large branching ratio ($\approx $ 50 \%), for identification of the $\phi$ meson in the phi production
experiments on cold nuclear matter. 
 
          Following the above method, in this paper we perform a detailed analysis
 of the $\phi$ production in $pA$ interactions at 2.83 GeV beam energy. This energy 
was employed in the recent measurements of proton-induced $\phi$ production in nuclei at the ANKE-COSY facility [15, 16]. 
We present the detailed predictions for the A- and momentum-dependences of the absolute and relative cross sections 
for the $\phi$ creation from these interactions
obtained in the framework of a nuclear spectral function approach [4, 17--21] for an incoherent primary proton--nucleon and secondary
pion--nucleon phi meson production processes in a different scenarios for the total $\phi$ in-medium width as well as with imposing the
ANKE spectrometer kinematical cuts on the laboratory $\phi$ momenta and production angles. In view of the future data from this
spectrometer, the predictions can be used as an important tool for extracting the valuable information on the phi in-medium properties.      
Some first results of this analysis have been reported in [22].

\section*{2 The model and inputs} 

\hspace{1.5cm} {\bf At first, we consider direct $\phi$ production mechanism.}
An incident proton can produce a $\phi$ directly in the first inelastic
$pN$ collision due to the nucleon's Fermi motion. 
Since we are interested in the near-threshold energy region, we have taken into account the following elementary processes 
which have the lowest free production threshold ($\approx$ 2.59 GeV):
\begin{equation}
p+p \to p+p+\phi,
\end{equation}
\begin{equation}
p+n \to p+n+\phi,
\end{equation}
\begin{equation}
p+n \to d+\phi.
\end{equation}
Because the phi--nucleon elastic cross section is rather small [1], we will neglect the elastic ${\phi}N$ final-state interactions
in the present study. Moreover, since the $\phi$ meson pole mass in the medium is approximately not affected by medium
effects [13, 14] as well as for reason of reducing the possible uncertainty of our calculations due to the use in them the model
nucleon self-energies, we will also ignore the medium modification of the outgoing hadron masses in the present work.
Then, taking into consideration the $\phi$ meson final-state absorption as well as assuming that the $\phi$ meson--as a narrow
resonance--is produced and propagated with its pole mass $M_{\phi}$ at small laboratory angles of our main interest, 
we can represent the inclusive cross section for the production on nuclei phi mesons with
the momentum ${\bf p}_{\phi}$ from the primary proton-induced reaction channels (1)--(3) as follows [22]:
\begin{equation}
\frac{d\sigma_{pA\to {\phi}X}^{({\rm prim})}
({\bf p}_0)}
{d{\bf p}_{\phi}}=I_{V}[A]\left[\frac{Z}{A}\left<\frac{d\sigma_{pp\to pp{\phi}}({\bf p}_{0}^{'},M_{\phi},
{\bf p}_{\phi})}{d{\bf p}_{\phi}}\right>+\frac{N}{A}\left<\frac{d\sigma_{pn\to pn{\phi}}({\bf p}_{0}^{'},M_{\phi},
{\bf p}_{\phi})}{d{\bf p}_{\phi}}\right>\right]+
\end{equation}
$$
+I_{V}[A]\frac{N}{A}\left<\frac{d\sigma_{pn\to d{\phi}}({\bf p}_{0}^{'},M_{\phi},
{\bf p}_{\phi})}{d{\bf p}_{\phi}}\right>,
$$
where
\begin{equation}
I_{V}[A]=2{\pi}A\int\limits_{0}^{R}r_{\bot}dr_{\bot}
\int\limits_{-\sqrt{R^2-r_{\bot}^2}}^{\sqrt{R^2-r_{\bot}^2}}dz
\rho(\sqrt{r_{\bot}^2+z^2})\times
\end{equation}
$$
\times    
\exp{\left[-\sigma_{pN}^{{\rm in}}A\int\limits_{-\sqrt{R^2-r_{\bot}^2}}^{z}
\rho(\sqrt{r_{\bot}^2+x^2})dx
-\int\limits_{z}^{\sqrt{R^2-r_{\bot}^2}}\frac{dx}
{\lambda_{\phi}(\sqrt{r_{\bot}^2+x^2},M_{\phi})}\right]},
$$
\begin{equation}
\lambda_{\phi}({\bf r},M_{\phi})=\frac{p_{\phi}}{M_{\phi}
\Gamma_{{\rm tot}}({\bf r},M_{\phi})}
\end{equation}
and
\begin{equation}
\left<\frac{d\sigma_{pN\to pN{\phi}(pn\to d{\phi})}({\bf p}_{0}^{'},M_{\phi},
{\bf p}_{\phi})}
{d{\bf p}_{\phi}}\right>=
\int\int 
P({\bf p}_t,E)d{\bf p}_tdE
\left[\frac{d\sigma_{pN\to pN{\phi}(pn\to d{\phi})}(\sqrt{s},M_{\phi},{\bf p}_{\phi})}
{d{\bf p}_{\phi}}\right].
\end{equation}
Here, 
$d\sigma_{pN\to pN{\phi}}(\sqrt{s},M_{\phi},{\bf p}_{\phi}) /d{\bf p}_{\phi}$ and 
$d\sigma_{pn\to d{\phi}}(\sqrt{s},M_{\phi},{\bf p}_{\phi}) /d{\bf p}_{\phi}$
are the off-shell
\footnote{The struck target nucleon is off-shell, see eq. (13) in [22].}
differential cross sections for $\phi$ production in reactions (1), (2) and (3), respectively, 
at the $pN$ center-of-mass energy $\sqrt{s}$; $\rho({\bf r})$ and 
$P({\bf p}_t,E)$ are the density and 
nuclear spectral function normalized to unity; 
${\bf p}_t$ and $E$ are the internal momentum and removal energy of the struck target nucleon 
just before the collision; $\sigma_{pN}^{{\rm in}}$ and $\Gamma_{{\rm tot}}({\bf r},M_{\phi})$  
are the inelastic cross section
of free $pN$ interaction and total $\phi$ width in its rest frame, taken at the point ${\bf r}$ inside the nucleus
and at the pole of the resonance;
$Z$ and $N$ are the numbers of protons and neutrons in 
the target nucleus ($A=N+Z$), $R$ is its radius;
${\bf p}_0$ and ${\bf p}_{0}^{'}$ are the momenta of the initial proton outside and inside the target nucleus.  

   The first term in eq. (4) describes the contribution to the $\phi$ meson production on nuclei from the primary $pN$ interactions
(1) and (2), whereas the second one represents the contribution to this production from the elementary process (3).

         Following [4, 23], we assume that the off-shell differential cross sections \\
$d\sigma_{pN\to pN{\phi}}(\sqrt{s},M_{\phi},{\bf p}_{\phi}) /d{\bf p}_{\phi}$ and 
$d\sigma_{pn\to d{\phi}}(\sqrt{s},M_{\phi},{\bf p}_{\phi}) /d{\bf p}_{\phi}$
for $\phi$ production in the reactions (1), (2) and (3), entering into eqs. (4), (7),  
are equivalent to the respective on-shell cross sections calculated for the off-shell kinematics of the elementary
processes  (1)--(3). In our approach the differential cross sections for $\phi$ production in the reactions (1), (2) have been
described [22] by the three-body phase space calculations normalized to the corresponding total cross sections
$\sigma_{pN \to pN{\phi}}(\sqrt{s})$. 

        For the free total cross section $\sigma_{pp \to pp{\phi}}(\sqrt{s})$ we have used the parametrization suggested in [22].
It fits quite well [22] the existing set of data [24, 25, 26] for the $pp \to pp{\phi}$ reaction in the threshold region.

     For obtaining the total cross section of the $pn \to pn{\phi}$ reaction, where data are not available, we have employed the
following relation:
\begin{equation}
  \sigma_{pn \to pn{\phi}}({\sqrt{s}})=f\left(\sqrt{s}-\sqrt{s_{th}}\right)\sigma_{pp \to pp{\phi}}({\sqrt{s}}).
\end{equation}
In the literature there are [27, 28, 29, 30] a different options to choose the experimentally unknown cross section ratio
$f=\sigma_{pn \to pn{\phi}}/{\sigma_{pp \to pp{\phi}}}$ in the near-threshold energy regime. 
We will adopt in our calculations for this ratio the excess energy $\epsilon$-dependence $f(\epsilon)$, obtained in [28]
within an effective meson--nucleon theory (see, also, [22]).

      Taking into consideration the two-body kinematics of the elementary process (3), we can readily get [22] the respective
expression for the differential cross section for $\phi$ production in this process.
This cross section is normalized to the corresponding total experimental cross section $\sigma_{pn \to d{\phi}}$ [27]. 
 In our calculations the angular distribution of  $\phi$ mesons in the $pn$ c.m.s. was assumed to be isotropic [27]. 

       For the $\phi$ production calculations in the case of  $^{12}$C and $^{27}$Al, $^{63}$Cu, $^{108}$Ag, $^{197}$Au, $^{238}$U 
target nuclei reported here we have employed for the nuclear density $\rho({\bf r})$,
respectively, the harmonic oscillator and the Woods-Saxon distributions, given in [22].
The nuclear spectral function $P({\bf p}_t,E)$ (which represents the
 probability to find a nucleon with momentum ${\bf p}_t$ and removal energy $E$ in the nucleus) for $^{12}$C target nucleus was taken
 from [17]. The single-particle part of this function for medium $^{27}$Al, $^{63}$Cu and heavy $^{108}$Ag, $^{197}$Au, $^{238}$U 
target nuclei was assumed to be the same as that for $^{208}$Pb [20, 21]. The latter was taken from [19]. The correlated part of the
nuclear spectral function for these target nuclei was borrowed from [17]. 

   Let us concentrate now on the total $\phi$ in-medium width appearing in (6) and used in the subsequent calculations of phi meson
attenuation in $pA$ interactions.

   For this width, the calculations foresee four different scenarios: {\bf i)} no in-medium effects and, correspondingly, the scenario with the
free $\phi$ width (dotted line in Fig. 1); {\bf ii)} the same in-medium effects on the masses of a daughter kaon, antikaon, $\rho$ meson and
through this on the $\phi$ decay width in a nuclear environment as well as the same collisional broadening
\footnote{Which is characterized by the collisional width of 10 MeV at saturation density $\rho_0$ [4].}
of a $\phi$ meson in this environment as those adopted before in [4] (solid line
\footnote{It shows that in this scenario the resulting total (decay+collisional) width of the $\phi$ meson grows as a function of the
density and reaches the value of around 30 MeV at the density $\rho_0$.}
in Fig. 1); {\bf iii)} the same in-medium effects on the phi decay width as those in the preceding case and the collisional broadening
of a $\phi$ meson, characterized by an enlarged phi collisional width of 25.4 MeV
 at normal nuclear matter density to get for the total $\phi$ width at this density the value of about 45 MeV predicted in ref. [14]
(dashed line in Fig. 1) and {\bf iv)} the same in-medium effects on the phi decay width as those in the previous cases
and the collisional broadening of a $\phi$ meson, determined by an increased phi collisional width of 40.4 MeV
at nuclear density $\rho_0$ to gain for the $\phi$ total width at this density the value of the order of 60 MeV, which is twice the
total width at $\rho_0$, used in the second scenario (dot-dashed line in Fig. 1). The parametrizations of [4] are taken here to calculate
the density dependences of the $\phi$ decay and collisional widths. 

{\bf Let us consider now the two-step $\phi$ production mechanism.}
At the bombarding energy of our interest (2.83 GeV) the following two-step $\phi$ production process with a pion
(which is assumed to be on-shell)
in an intermediate state may contribute to the phi production in $pA$ interactions:
\begin{equation}
p+N_1 \to \pi+X,
\end{equation}
\begin{equation}
\pi+N_2 \to \phi+N,
\end{equation}
provided that the latter subprocess is allowed energetically.
Taking into account the phi final-state absorption and ignoring the influence 
\footnote{In line with the assumption about the absence of such influence on the final hadron masses in primary proton-induced
reaction channels (1)--(3).}
of the nuclear environment
on the outgoing hadron masses in the $\phi$ production channel (10), we get the
following expression for the phi production cross section for $pA$ reactions at small laboratory angles of interest from this
channel [22]:
\begin{equation}
\frac{d\sigma_{pA\to {\phi}X}^{({\rm sec})}
({\bf p}_0)}
{d{\bf p}_{\phi}}=\frac{I_{V}^{({\rm sec})}[A]}{I_{V}^{'}[A]}
\sum_{\pi=\pi^+,\pi^0,\pi^-}\int \limits_{4\pi}d{\bf \Omega}_{\pi}
\int \limits_{p_{\pi}^{{\rm abs}}}^{p_{\pi}^{{\rm lim}}
(\vartheta_{\pi})}p_{\pi}^{2}
dp_{\pi}
\frac{d\sigma_{pA\to {\pi}X}^{({\rm prim})}({\bf p}_0)}{d{\bf p}_{\pi}}\times
\end{equation}
$$
\times
\left[\frac{Z}{A}\left<\frac{d\sigma_{{\pi}p\to{\phi}N}({\bf p}_{\pi},
{\bf p}_{\phi})}{d{\bf p}_{\phi}}\right>+\frac{N}{A}\left<\frac{d\sigma_{{\pi}n\to{\phi}N}({\bf p}_{\pi},
{\bf p}_{\phi})}{d{\bf p}_{\phi}}\right>\right],
$$
where
\begin{equation}
I_{V}^{({\rm sec})}[A]=2{\pi}A^2\int\limits_{0}^{R}r_{\bot}dr_{\bot}
\int\limits_{-\sqrt{R^2-r_{\bot}^2}}^{\sqrt{R^2-r_{\bot}^2}}dz
\rho(\sqrt{r_{\bot}^2+z^2})
\int\limits_{0}^{\sqrt{R^2-r_{\bot}^2}-z}dl
\rho(\sqrt{r_{\bot}^2+(z+l)^2})
\times
\end{equation}
$$
\times    
\exp{\left[-\sigma_{pN}^{{\rm in}}A\int\limits_{-\sqrt{R^2-r_{\bot}^2}}^{z}
\rho(\sqrt{r_{\bot}^2+x^2})dx
-\sigma_{{\pi}N}^{{\rm tot}}A\int\limits_{z}^{z+l}
\rho(\sqrt{r_{\bot}^2+x^2})dx\right]}
\times
$$
$$
\times
\exp{\left[-\int\limits_{z+l}^{\sqrt{R^2-r_{\bot}^2}}\frac{dx}
{\lambda_{\phi}(\sqrt{r_{\bot}^2+x^2},M_{\phi})}\right]},
$$
\begin{equation}
I_{V}^{'}[A]=2{\pi}A\int\limits_{0}^{R}r_{\bot}dr_{\bot}
\int\limits_{-\sqrt{R^2-r_{\bot}^2}}^{\sqrt{R^2-r_{\bot}^2}}dz
\rho(\sqrt{r_{\bot}^2+z^2})\times
\end{equation}
$$
\times    
\exp{\left[-\sigma_{pN}^{{\rm in}}A\int\limits_{-\sqrt{R^2-r_{\bot}^2}}^{z}
\rho(\sqrt{r_{\bot}^2+x^2})dx
-\sigma_{{\pi}N}^{{\rm tot}}A\int\limits_{z}^{\sqrt{R^2-r_{\bot}^2}}
\rho(\sqrt{r_{\bot}^2+x^2})dx\right]},
$$
\begin{equation}
\left<\frac{d\sigma_{{\pi}N\to {\phi}N}({\bf p}_{\pi},
{\bf p}_{\phi})}
{d{\bf p}_{\phi}}\right>=
\int\int 
P({\bf p}_t,E)d{\bf p}_tdE
\left[\frac{d\sigma_{{\pi}N\to {\phi}N}(\sqrt{s_1},{\bf p}_{\phi})}
{d{\bf p}_{\phi}}\right],
\end{equation}
\begin{equation}                                 
\cos{\vartheta_{\pi}}={\bf \Omega}_0{\bf \Omega}_{\pi},\,\,\,\,
{\bf \Omega}_{0}={\bf p}_{0}/p_{0},\,\,\,\,{\bf \Omega}_{\pi}={\bf p}_{\pi}/p_{\pi}.
\end{equation}
Here, $d\sigma_{pA\to {\pi}X}^{({\rm prim})}({\bf p}_0)/d{\bf p}_{\pi}$ are the
inclusive differential cross sections for pion production on nuclei at small laboratory angles and for high momenta from
the primary proton-induced reaction channel (9); $d\sigma_{{\pi}N\to {\phi}N}(\sqrt{s_1},{\bf p}_{\phi})/d{\bf p}_{\phi}$ is
the free inclusive differential cross section for $\phi$ production via the subprocess (10) calculated for the off-shell kinematics of
this subprocess at the ${\pi}N$ center-of-mass energy $\sqrt{s_1}$;
$\sigma_{\pi N}^{{\rm tot}}$ is the total cross section of the free $\pi N$ interaction;
${\bf p}_{\pi}$ is the momentum of a pion;
$p_{\pi}^{{\rm abs}}$ is the absolute threshold momentum for phi production on the residual nucleus by an intermediate pion;
$p_{\pi}^{{\rm lim}}(\vartheta_{\pi})$ is the kinematical limit for pion production
at the lab angle $\vartheta_{\pi}$ from proton-nucleus collisions. The quantity $\lambda_{\phi}$ is defined above by eq. (6).

    The expression for the differential cross section for $\phi$ production in the elementary process (10) has the form similar to
that for proton-induced reaction (3) [22]. The $\phi$ angular distribution in the ${\pi}N$ c.m.s., entering into it, 
was assumed to be isotropic [1] in our calculations of phi creation in $pA$ collisions from reaction (10).
The elementary $\phi$ production reactions  ${\pi}^+n \to {\phi}p$, ${\pi}^0p \to {\phi}p$, ${\pi}^0n \to {\phi}n$ and
${\pi}^-p \to {\phi}n$ have been included in our calculations of the $\phi$ production on nuclei.
The total cross sections of the first three reactions are related by the isospin considerations to 
the total cross section of the last reaction [22].
For the free total cross section of the reaction ${\pi}^{-}p \to {\phi}n$ we have used the parametrization suggested in [31].

   Another a very important ingredients for the calculation of the phi production cross section in proton--nucleus reactions
from pion-induced reaction channel (10)--the high momentum parts of the differential cross sections for pion production on
nuclei at small lab angles from the primary process (9)--for $^{63}$Cu target nucleus were taken from [20]. For $\phi$
production calculations in the case of $^{12}$C and $^{27}$Al, $^{108}$Ag, $^{197}$Au, $^{238}$U target nuclei presented
below we have supposed [21] that the ratio of the differential cross section for pion creation on $^{12}$C and on these nuclei
from the primary process (9) to the effective number
\footnote{Which is given by eq. (13).}
of nucleons participating in it is the same as that for $^{9}$Be and $^{63}$Cu adjusted for the kinematics relating, respectively, to
$^{12}$C and $^{27}$Al, $^{108}$Ag, $^{197}$Au, $^{238}$U.   

  Now, let us proceed to the discussion of the results of our calculations for phi production in $pA$ interactions in the
framework of the model outlined above.

\section*{3 Results}

\hspace{1.5cm}The authors of ref. [3] have suggested to use as a measure for the $\phi$ meson width in nuclei the following relative
observable--the double ratio: $R(^{A}X)/R(^{12}{\rm C})=
(\sigma_{pA \to {\phi}X}/A)/( \sigma_{p^{12}{\rm C} \to {\phi}X}/12)$,
i.e. the ratio of the nuclear total phi production cross section from $pA$ reactions divided by A to the same quantity on $^{12}$C.
But instead of this ratio, first we consider the following analogous ratio
$R(^{A}X)/R(^{12}{\rm C})=\\
({\tilde \sigma}_{pA \to {\phi}X}/A)/({\tilde \sigma}_{p^{12}{\rm C} \to {\phi}X}/12)$, where
${\tilde \sigma}_{pA \to {\phi}X}$ is the nuclear cross section for $\phi$ production from proton--nucleus collisions in the whole
ANKE acceptance window:\\
 $0.6~{\rm {GeV/c}} \le p_{\phi} \le 1.8~{\rm {GeV/c}}$ and $0^{\circ} \le \theta_{\phi} \le 8^{\circ}$.
It is clear that our predictions for the latter ratio could be directly compared with the future data from the ANKE experiment [15, 16]
to extract the definite information on the $\phi$ width in the nuclear matter.

      Figure 2 shows the ratio
$R(^{A}X)/R(^{12}{\rm C})$=$({\tilde \sigma}_{pA \to {\phi}X}/A)/({\tilde \sigma}_{p^{12}{\rm C} \to {\phi}X}/12)$
as a function of the nuclear mass number A calculated on the basis of eqs. (4), (11)
for the one-step and one- plus two-step
$\phi$ creation mechanisms (corresponding lines with symbols "prim" and "prim+sec" by them) for the projectile
energy of 2.83 GeV as well as for the cross section ratio $\sigma_{pn \to pn{\phi}}/{\sigma_{pp \to pp{\phi}}}$ 
in the excess-energy-dependent form $f(\epsilon)$ from [28]
and within
the first two scenarios for the $\phi$ in-medium width: {\bf i)} free phi width (dotted lines), 
{\bf ii)} in-medium phi width shown by solid curve in Fig. 1 (solid lines).  
It can be seen that there are clear differences between the results obtained by using, on the one hand, different $\phi$ in-medium
widths under consideration and the same assumptions concerning the phi production mechanism (between corresponding dotted
and solid lines), and, on the other hand, different suppositions about the $\phi$ creation mechanism and the same phi widths in the
medium (between dotted lines, and between solid lines). 
We may see, for example, that in the chosen kinematics for heavy nuclei, where the $\phi$
absorption and $\phi$ creation via the secondary pion-induced reaction channel ${\pi}N \to {\phi}N$ are enhanced, 
the calculated ratio $R(^{A}X)/R(^{12}{\rm C})$ can be of the order of 0.30 and 0.45 for the direct $\phi$
production mechanism as well as 0.35 and 0.52 for the direct plus two-step phi creation mechanisms in the cases when the absorption
of phi mesons in nuclear matter was determined, respectively, by their in-medium width shown by solid curve in Fig. 1 and by their free
width. Therefore, we can conclude that the observation of the A-dependence, like that just considered, can serve as an important
tool to determine the $\phi$ width in nuclei and in the analysis of the observed in the whole ANKE acceptance window 
dependence it is needed to account for the secondary pion-induced phi production processes.

     In Fig. 3 we show the predictions of our model for the double ratio $R(^{A}X)/R(^{12}{\rm C})$ of interest obtained for
initial energy of 2.83 GeV by considering primary proton--nucleon (1)--(3) and secondary pion--nucleon (10) $\phi$ production
processes as well as by using for the cross section ratio $\sigma_{pn \to pn{\phi}}/{\sigma_{pp \to pp{\phi}}}$
the excess-energy-dependent form $f(\epsilon)$ from [28]
 and for the $\phi$ in-medium width those shown by dashed and dot-dashed curves in Fig. 1
(respectively, dashed and dot-dashed lines in Fig. 3) in comparison to the previously performed calculations with employing for this
width the free one (dotted line in Fig. 3) and that depicted by solid curve in Fig. 1 (solid line in Fig. 3). We observe in this figure the
visible differences between all calculations corresponding to different scenarios for the $\phi$ width in nuclei, which means that the
future data from the ANKE-at-COSY experiment [15, 16] should help, as one may hope, to distinguish between these scenarios.
   
      In the next step, we consider the double ratio $R(^{A}X)/R(^{12}{\rm C})$, as defined above, in the $\phi$ low-momentum
$0.6~{\rm {GeV/c}} \le p_{\phi} \le 0.9~{\rm {GeV/c}}$ and high-momentum $1.5~{\rm {GeV/c}} \le p_{\phi} \le 1.8~{\rm {GeV/c}}$
intervals belonging to the acceptance window of the ANKE  spectrometer to investigate the relative role of the primary and
secondary $\phi$ meson production processes at different phi momenta in $pA$ reactions at considered incident energy. 
The corresponding ratios are shown in Figs. 4 and 5. Looking at these figures, one can see that the secondary pion-induced
reaction channel ${\pi}N \to {\phi}N$ plays a minor role for all considered target nuclei in the phi 
high-momentum interval $1.5~{\rm {GeV/c}} \le p_{\phi} \le 1.8~{\rm {GeV/c}}$. This observation is very important, since it
enables us to put a strong constraints on the in-medium $\phi$ width from comparison the results of more reliable model
calculations, based only on the direct phi production mechanism, with the respective high-momentum data from the ANKE
experiment [15, 16].

   In Fig. 6 we show together the results of our calculations, given before separately in Figs. 4, 5, for the A-dependence of present
interest for the primary (1)--(3) plus secondary (10) $\phi$ production processes obtained for bombarding energy of 2.83 GeV by
employing in them both the free phi mesons width (dotted lines) and their in-medium width shown by solid curve in Fig. 1 (solid lines)
as well as the considered options
\footnote{Indicated by the respective symbols in the figure and by the lines.}
for the ratio $\sigma_{pn \to pn{\phi}}/{\sigma_{pp \to pp{\phi}}}$ and for the $\phi$ momentum bin 
to see also the sensitivity of the calculated A-dependence to the phi momentum. 
One can see that the differences between the calculations for the same $\phi$ width shown by solid curve in Fig. 1 with adopting
for the $\phi$ momentum bin two options: $0.6~{\rm {GeV/c}} \le p_{\phi} \le 0.9~{\rm {GeV/c}}$ and
$1.5~{\rm {GeV/c}} \le p_{\phi} \le 1.8~{\rm {GeV/c}}$ (between solid curves) are insignificant, which means
that this relative observable depends weakly on the $\phi$ momentum and measurement of such weak dependence should,
in particular, reflect the fact that the phi width in the medium is momentum-independent. 

           In Fig. 7 we show the predictions of our model for the ratio $R(^{A}X)/R(^{12}{\rm C})$ of interest obtained for
beam energy of 2.83 GeV by considering primary proton--nucleon (1)--(3) and secondary pion--nucleon (10) $\phi$ production
processes, the $\phi$ low-momentum bin $0.6~{\rm {GeV/c}} \le p_{\phi} \le 0.9~{\rm {GeV/c}}$ as well as by using for the phi
in-medium width those shown by dashed and dot-dashed curves in Fig. 1
(respectively, dashed and dot-dashed lines in Fig. 7) in comparison to the before performed calculations (see Fig. 4)
with employing for this width the free one (dotted line in Fig. 7) and that depicted by solid curve in Fig. 1 (solid line in Fig. 7).
Comparing the dashed and dot-dashed lines in Figs. 5 and 7, corresponding to the calculations of the ratio
$R(^{A}X)/R(^{12}{\rm C})$ in the $\phi$ high- and low-momentum bins as well as in the last two adopted scenarios for the
phi width in the medium, characterized by an enlarged $\phi$ total width of 45 and 60 MeV at the density $\rho_0$, one can see
that this relative observable, as in the preceding case, depends weakly on the $\phi$ momentum also in the case of yet increased
phi in-medium total width.  

        Finally, in Fig. 8 we show the predictions of our model for the momentum-dependence of the absolute $\phi$ meson yield
from the one- plus two-step phi creation mechanisms in $p$$^{197}$Au collisions obtained for proton kinetic energy of 2.83 GeV
in employed four scenarios for the total $\phi$ in-medium width as well as with imposing the ANKE spectrometer kinematical cuts 
on the laboratory phi momenta and production angles. By looking at this figure, we see that the absolute $\phi$ meson yield,
along with the relative one considered before, is appreciably sensitive to its in-medium width and the measurements of such
sensitivity are favourable at high phi momenta where the cross section for phi production is larger.

    Thus, our results demonstrate that the measurements of the A- and momentum-dependences of
the absolute and relative cross sections for phi production in $pA$ reactions in the considered kinematics 
and at above threshold beam energies 
will allow indeed to shed light on the $\phi$ in-medium properties. They show also that to extract the definite information on the 
$\phi$ width in nuclear matter from the analysis of the measured such dependences it is needed to take into account in this analysis
the secondary pion-induced phi production processes.

\section*{4 Conclusions}

\hspace{1.5cm} In this paper we have calculated the A- and momentum-dependences of the absolute and relative cross sections 
for $\phi$ production
from $pA$ reactions at 2.83 GeV beam energy by considering incoherent primary proton--nucleon and secondary pion--nucleon
phi production processes in the framework of a nuclear spectral function approach [4, 17--22], which takes properly into account
the struck target nucleon momentum and removal energy distribution, novel elementary cross sections for proton--nucleon 
reaction channel close to threshold as well as different scenarios for the total $\phi$ width in the medium
and the ANKE spectrometer kinematical cuts on the laboratory phi momenta and production angles. 
It was found that the secondary pion-induced reaction channel ${\pi}N \to {\phi}N$ contributes
distinctly to the "low-momentum" $\phi$ production in heavy nuclei in the chosen kinematics 
and, hence, it has to be taken into consideration on close examination of these dependences with the aim to get information
on the phi width in the nuclear matter. It was also shown that both
the A-dependence of the relative $\phi$ meson production cross section and momentum-dependence of the absolute $\phi$ 
meson yield from $pA$ collisions in the considered
kinematics and at incident energy of interest are appreciably sensitive to the absorption of the $\phi$ in the
surrounding nuclear matter in its way out of the nucleus, which is governed in turn by its in-medium width. This gives a nice 
opportunity to obtain the definite information on the $\phi$ width in the nuclear medium at finite baryon densities from the direct
comparison the results of our present calculations for these observables with the future data from the recently performed
ANKE-at-COSY experiment [15, 16].

            The author gratefully acknowledges stimulating discussions with M. Hartmann, Yu. Kiselev, V. Koptev, A. Sibirtsev, 
H. Str$\ddot{\rm o}$her. This work is partly supported by the Russian Fund for Basic Research, Grant No.07-02-91565.

\newpage
\begin{figure}[h!]
\centerline{\epsfig{file=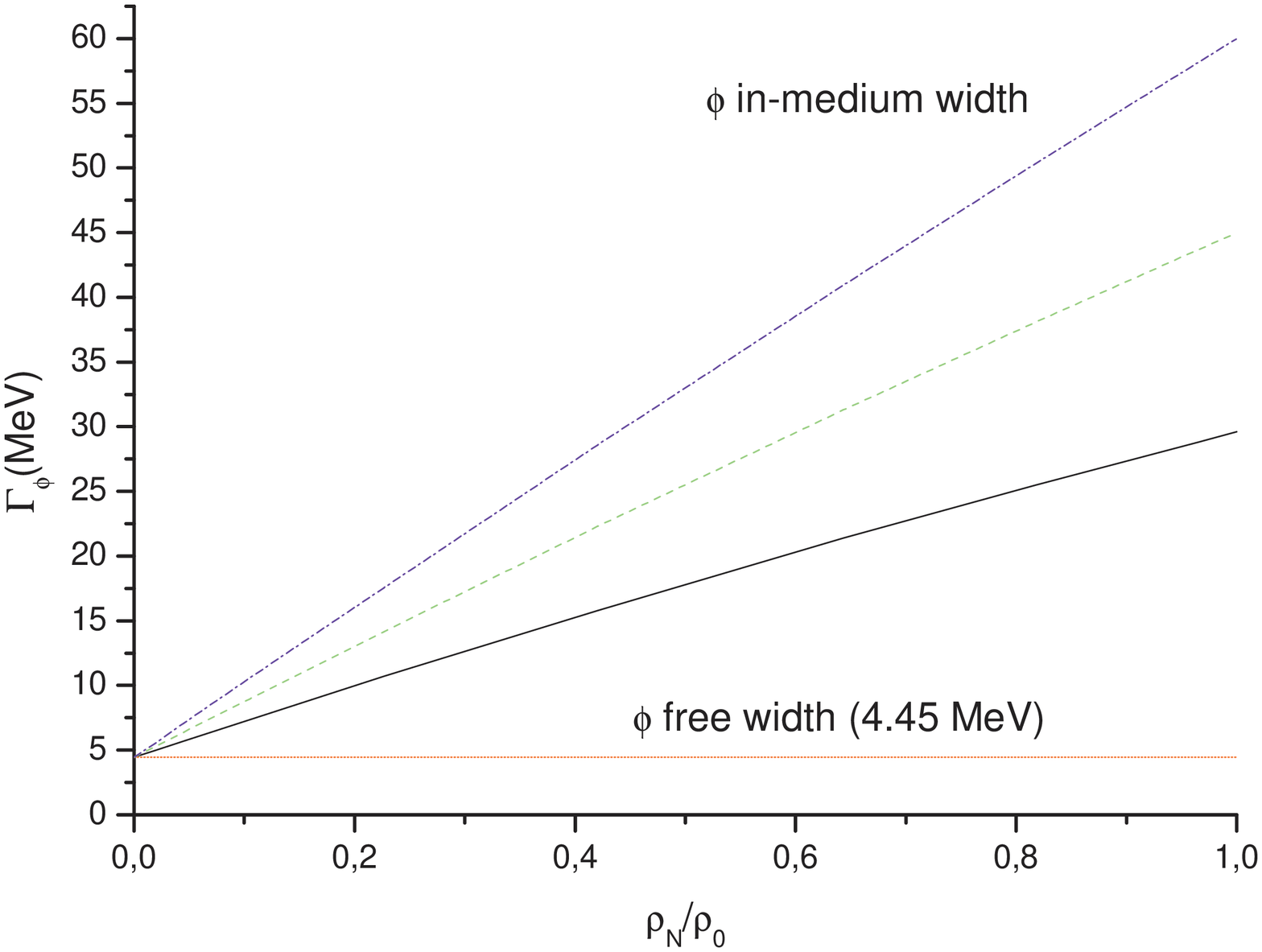,width=.88\textwidth,angle=270,silent=,
clip=}}
\caption{\label{centered}
Phi meson total width as a function of density. For notation see the 
text.}
\end{figure}
\begin{figure}[h!]
\centerline{\epsfig{file=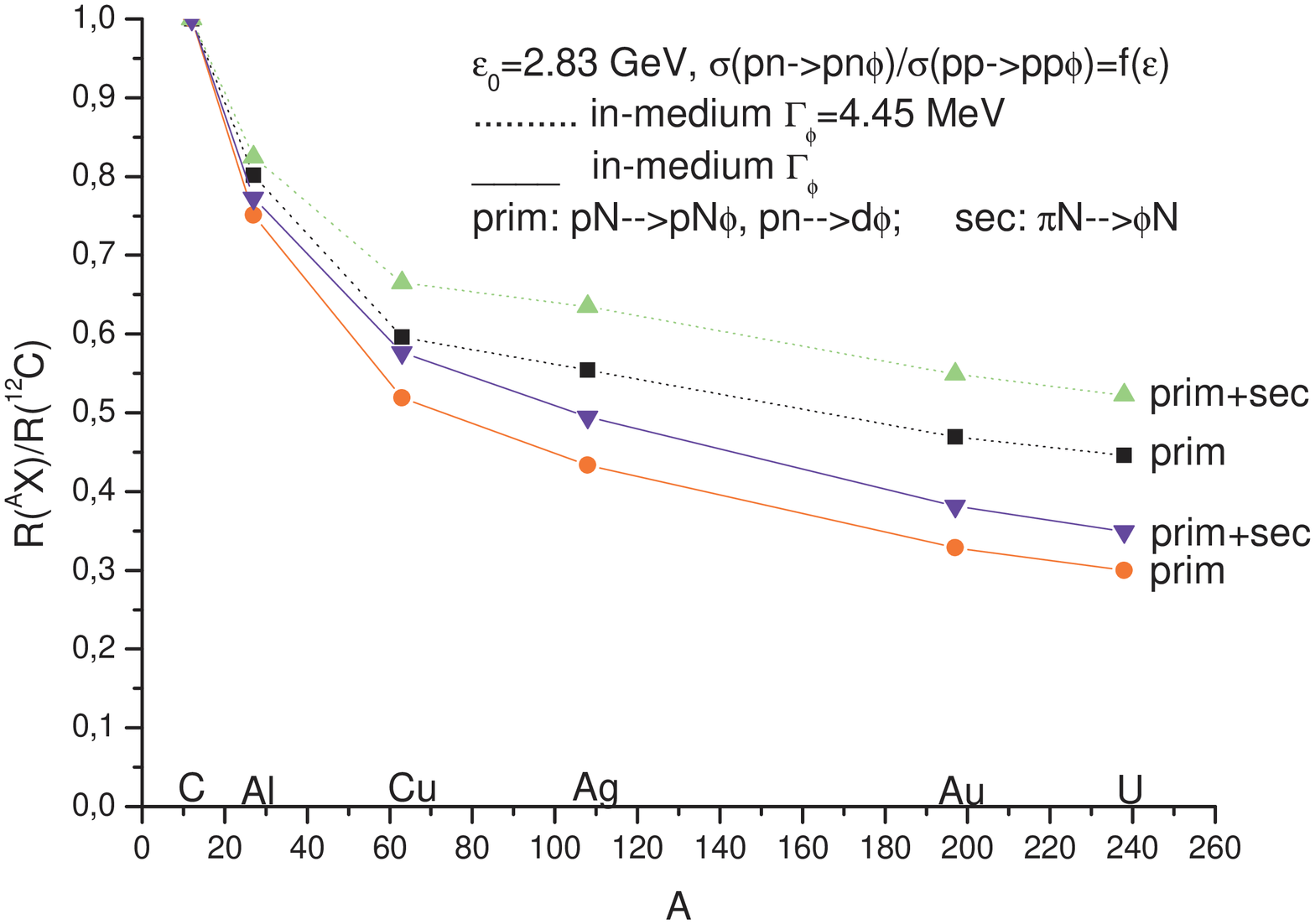,width=.88\textwidth,angle=270,silent=,
clip=}}
\caption{\label{centered}
Ratio $R(^{A}X)/R(^{12}{\rm C})$=
$({\tilde \sigma}_{pA \to {\phi}X}/A)/({\tilde \sigma}_{p^{12}{\rm C} \to {\phi}X}/12)$ as a function of the nuclear mass
number for initial energy of 2.83 GeV and for the cross section ratio $\sigma_{pn \to pn{\phi}}/{\sigma_{pp \to pp{\phi}}}$
in the excess-energy-dependent form $f(\epsilon)$ from [28]
calculated within the different scenarios for the $\phi$ meson production mechanism and for its in-medium width. For notation see
the text.}
\end{figure}
\begin{figure}[h!]
\centerline{\epsfig{file=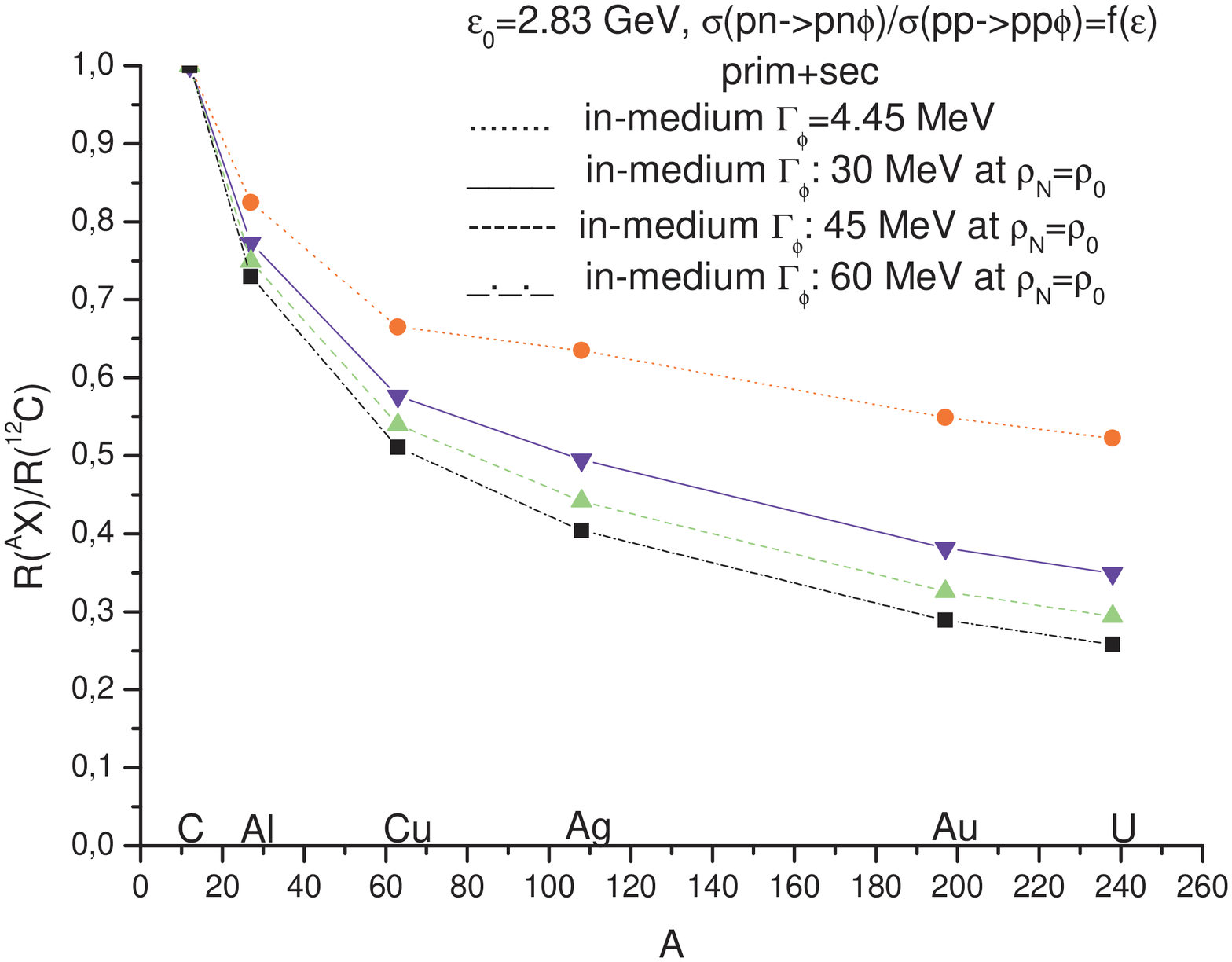,width=.88\textwidth,angle=270,silent=,
clip=}}
\caption{\label{centered}
Ratio $R(^{A}X)/R(^{12}{\rm C})$=
$({\tilde \sigma}_{pA \to {\phi}X}/A)/({\tilde \sigma}_{p^{12}{\rm C} \to {\phi}X}/12)$ as a function of the nuclear mass
number for the one- plus two-step $\phi$ production mechanisms calculated for incident energy of 2.83 GeV and for
the cross section ratio $\sigma_{pn \to pn{\phi}}/{\sigma_{pp \to pp{\phi}}}$ in the excess-energy-dependent form 
$f(\epsilon)$ from [28] within the different scenarios for the phi 
in-medium width. For notation see the text.}
\end{figure}
\begin{figure}[h!]
\centerline{\epsfig{file=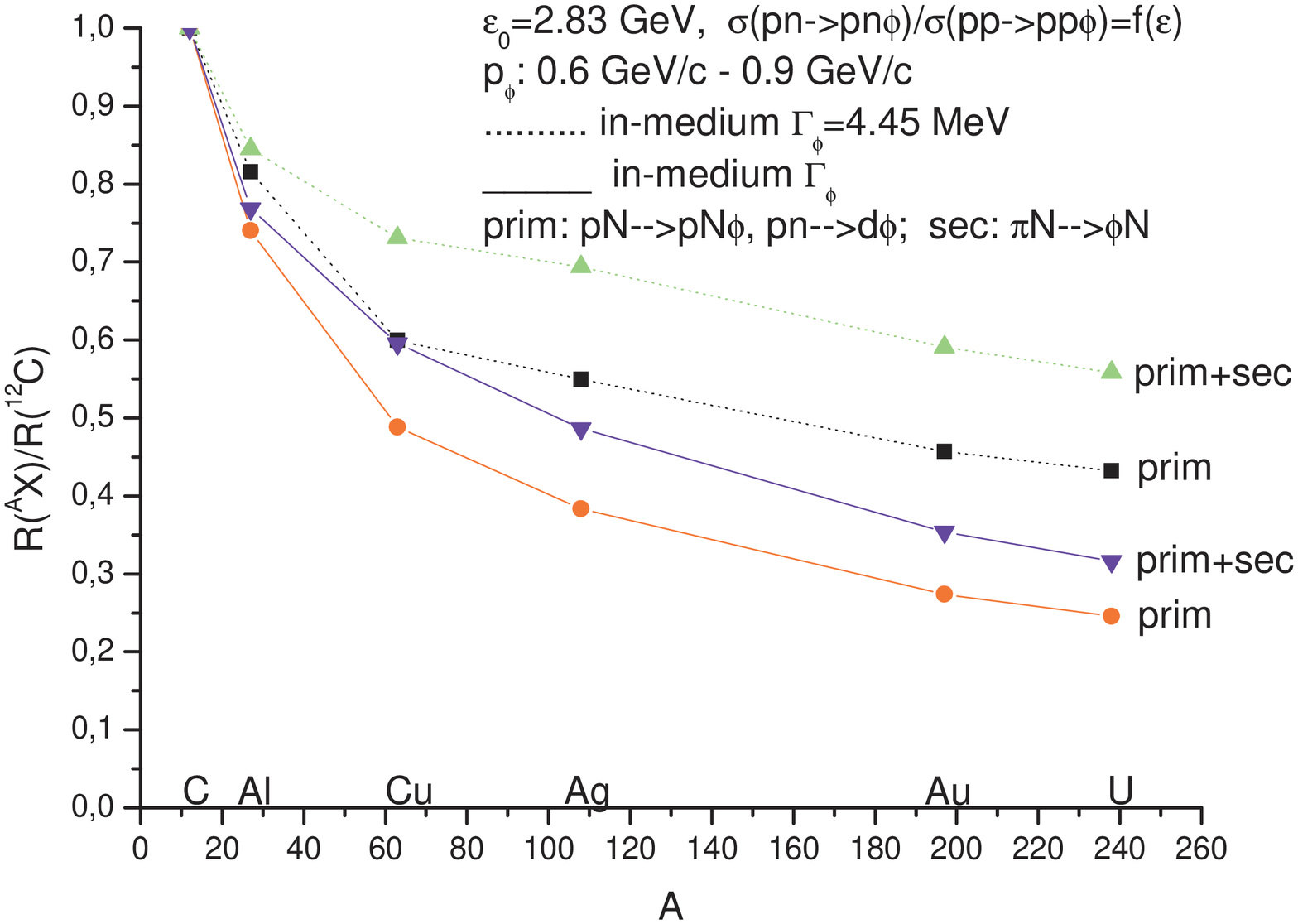,width=.88\textwidth,angle=270,silent=,
clip=}}
\caption{\label{centered}
The same as in Fig. 2, but it is supposed that the nuclear cross section ${\tilde \sigma}_{pA \to {\phi}X}$
for $\phi$ production from $pA$ collisions is calculated
in the phi momentum range $0.6~{\rm {GeV/c}} \le p_{\phi} \le 0.9~{\rm {GeV/c}}$ and in the phi polar angle
domain $0^{\circ} \le \theta_{\phi} \le 8^{\circ}$ in the lab system.}
\end{figure}
\begin{figure}[h!]
\centerline{\epsfig{file=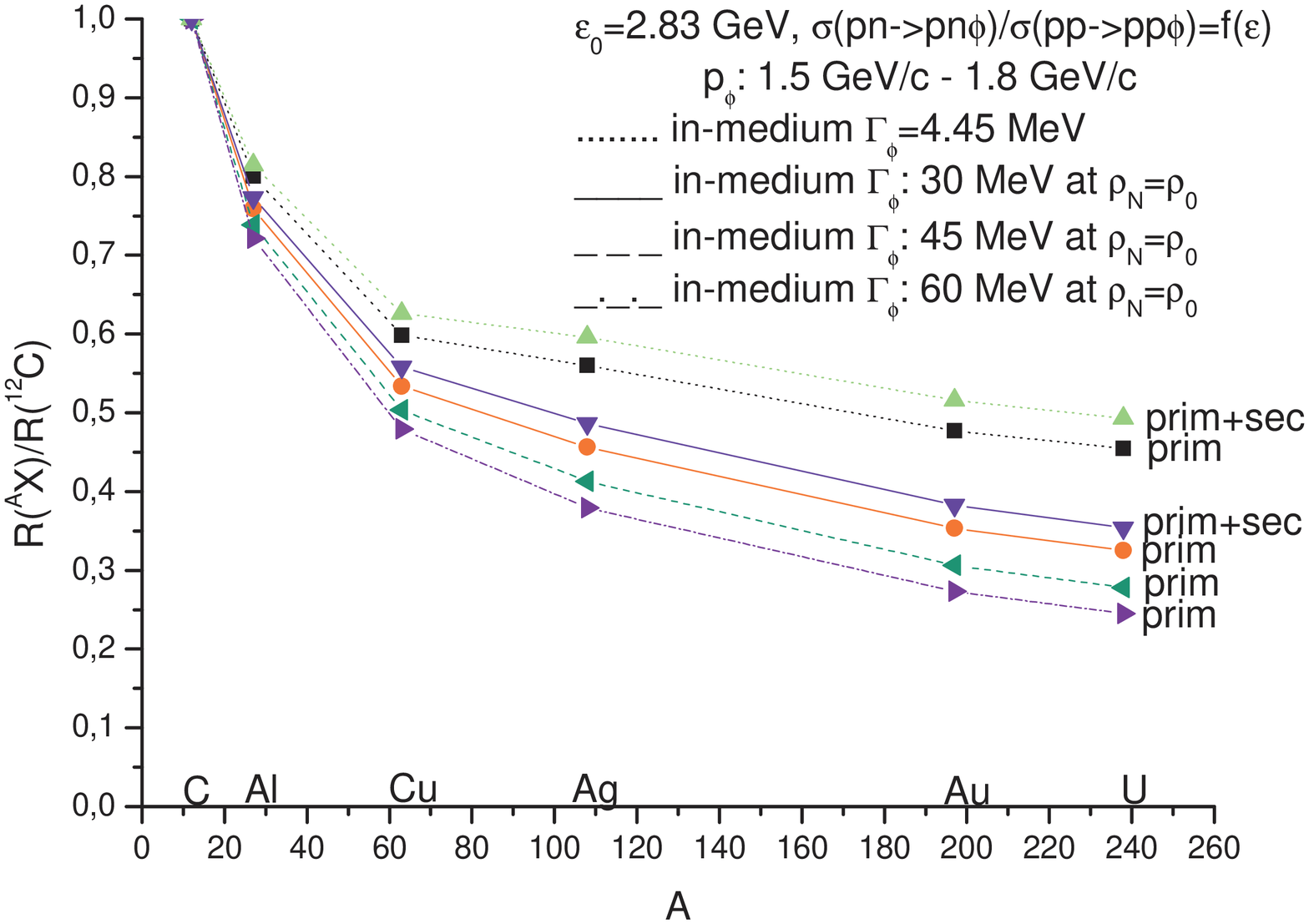,width=.88\textwidth,angle=270,silent=,
clip=}}
\caption{\label{centered}
Ratio $R(^{A}X)/R(^{12}{\rm C})$=
$({\tilde \sigma}_{pA \to {\phi}X}/A)/({\tilde \sigma}_{p^{12}{\rm C} \to {\phi}X}/12)$ as a function of the nuclear mass
number calculated for the one-step and one- plus two-step $\phi$ production mechanisms (corresponding lines with symbols
"prim" and "prim+sec" by them) for incident energy of 2.83 GeV as well as for 
the cross section ratio $\sigma_{pn \to pn{\phi}}/{\sigma_{pp \to pp{\phi}}}$ in the excess-energy-dependent form 
$f(\epsilon)$ from [28] and within the four scenarios for the phi in-medium width, considered in Fig. 1 and indicated by the values
of the total $\phi$ in-medium width at saturation density $\rho_0$. The nuclear cross section ${\tilde \sigma}_{pA \to {\phi}X}$
for $\phi$ production from $pA$ collisions is calculated
in the phi momentum range $1.5~{\rm {GeV/c}} \le p_{\phi} \le 1.8~{\rm {GeV/c}}$ and in the phi polar angle
domain $0^{\circ} \le \theta_{\phi} \le 8^{\circ}$ in the lab system.}
\end{figure}
\begin{figure}[h!]
\centerline{\epsfig{file=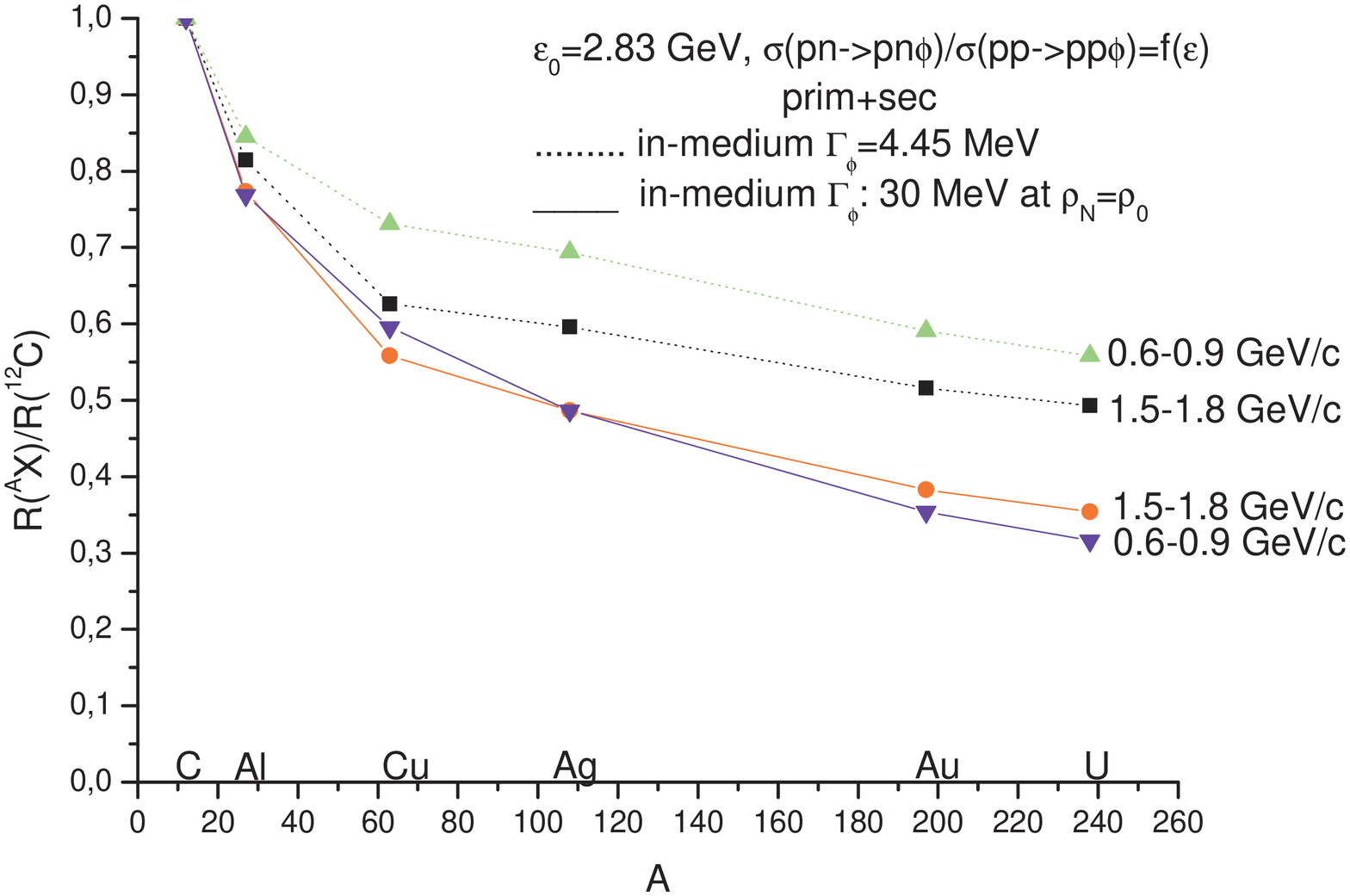,width=.88\textwidth,angle=270,silent=,
clip=}}
\caption{\label{centered}
Ratio $R(^{A}X)/R(^{12}{\rm C})$=
$({\tilde \sigma}_{pA \to {\phi}X}/A)/({\tilde \sigma}_{p^{12}{\rm C} \to {\phi}X}/12)$ as a function of the nuclear mass
number for the one- plus two-step $\phi$ production mechanisms calculated for incident energy of 2.83 GeV within the different
scenarios for the phi in-medium width and for 
the choice of the $\phi$ momentum bin. For notation see the text.}
\end{figure}
\begin{figure}[h!]
\centerline{\epsfig{file=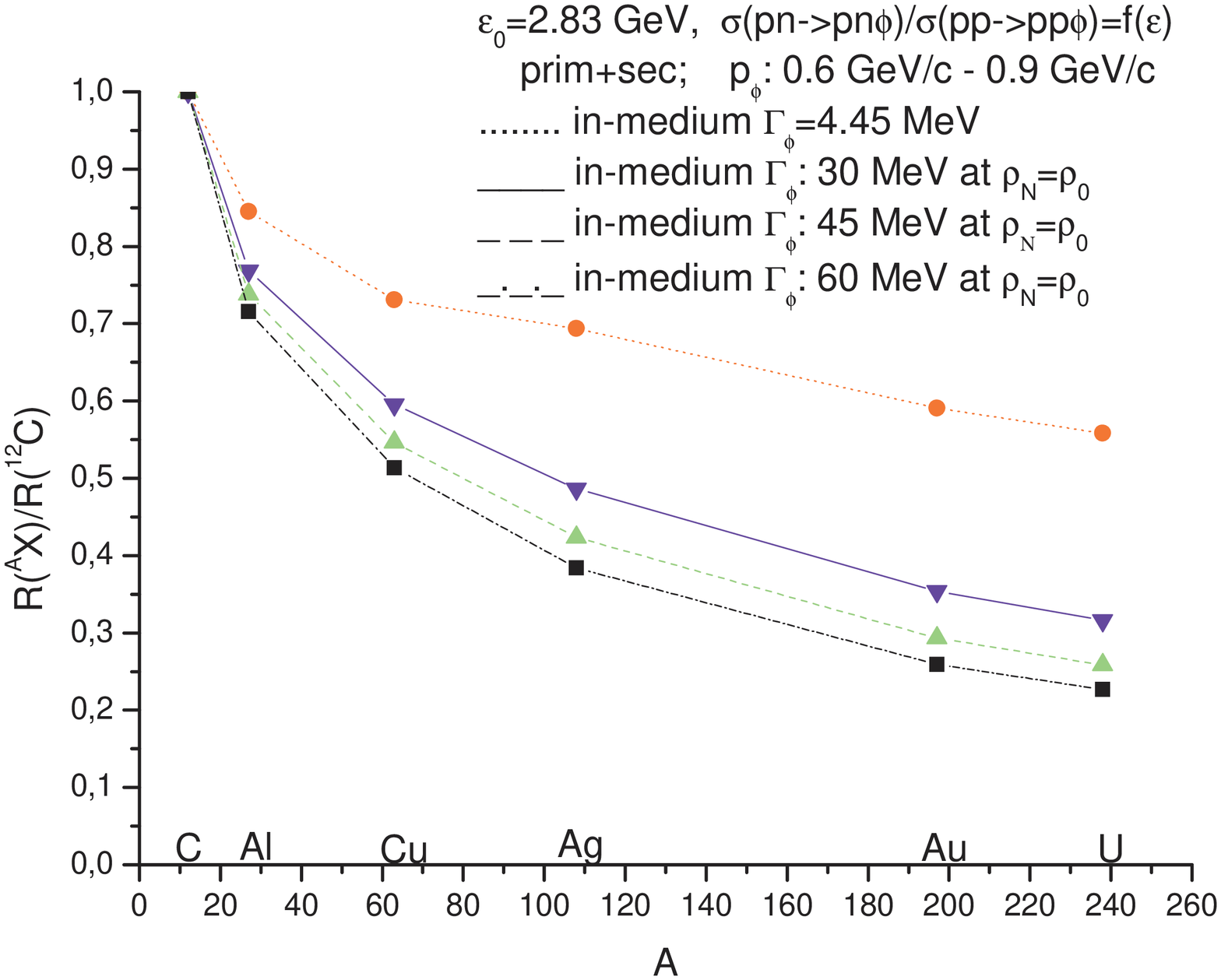,width=.88\textwidth,angle=270,silent=,
clip=}}
\caption{\label{centered}
Ratio $R(^{A}X)/R(^{12}{\rm C})$=
$({\tilde \sigma}_{pA \to {\phi}X}/A)/({\tilde \sigma}_{p^{12}{\rm C} \to {\phi}X}/12)$ as a function of the nuclear mass
number for the one- plus two-step $\phi$ production mechanisms calculated for incident energy of 2.83 GeV  
within the different scenarios for the phi in-medium width. 
For notation see the text.}
\end{figure}
\begin{figure}[h!]
\centerline{\epsfig{file=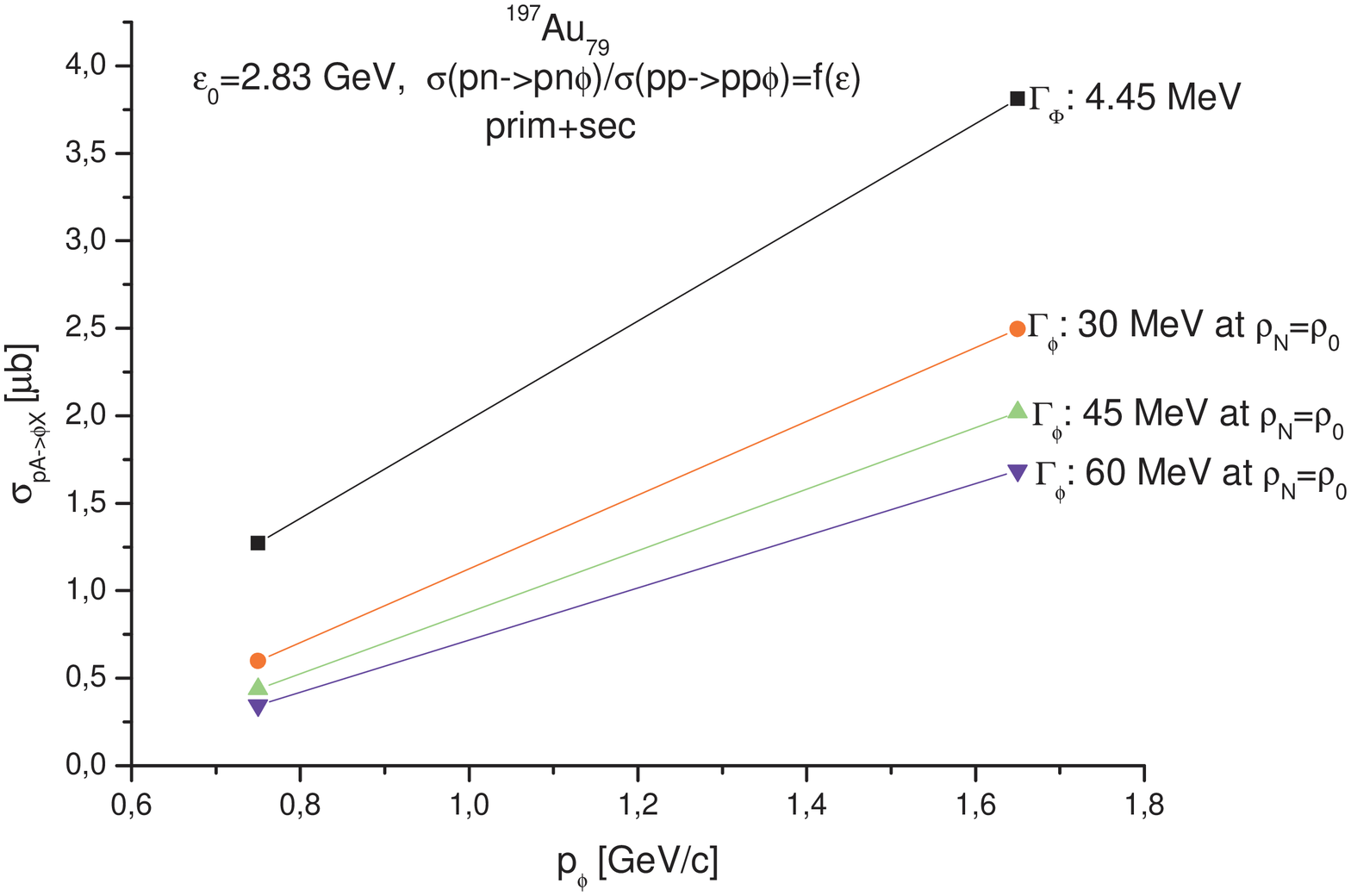,width=.88\textwidth,angle=270,silent=,
clip=}}
\caption{\label{centered}
The total cross section of $\phi$ production by 2.83 GeV protons from primary (1)--(3) and secondary (10)
channels in $p$$^{197}$Au collisions in adopted scenarios 
for the total $\phi$ in-medium width (indicated by the values of this width at density $\rho_0$ by the lines)
in the phi polar angle domain $0^{\circ} \le \theta_{\phi} \le 8^{\circ}$ and
in the phi momentum ranges $0.6~{\rm {GeV/c}} \le p_{\phi} \le 0.9~{\rm {GeV/c}}$,
$1.5~{\rm {GeV/c}} \le p_{\phi} \le 1.8~{\rm {GeV/c}}$ in the lab system (respectively, left and right points in the figure).
The lines are included to guide the eyes.}
\end{figure}
\end{document}